\documentclass[b5paper,twoside,reqno]{bjp}
\usepackage{cite}
\usepackage{latexsym}
\usepackage{graphicx}
\usepackage{amssymb,amsmath,amsfonts,amsbsy, amsthm, xspace, latexsym, amscd, enumitem}
\usepackage[utf8]{inputenc}
\usepackage[T1]{fontenc}
\usepackage[english]{babel}
\usepackage{times}
\usepackage{url,bm}
\newcommand\rf[1]{(\ref{eq:#1})}
\newcommand\lab[1]{\label{eq:#1}}
\newcommand\nonu{\nonumber}
\newcommand\br{\begin{eqnarray}}
\newcommand\er{\end{eqnarray}}
\newcommand\be{\begin{equation}}
\newcommand\ee{\end{equation}}

\newcommand\lb{\lbrack}
\newcommand\rb{\rbrack}

\renewcommand\({\left(}
\renewcommand\){\right)}

\newcommand\bc{\begin{center}}
\newcommand\ec{\end{center}}

\newcommand\partder[2]{\frac{{\partial {#1}}}{{\partial {#2}}}}

\renewcommand\a{\alpha}

\newcommand\eps{\epsilon}
\newcommand\vareps{\varepsilon}

\newcommand\G{\Gamma}

\newcommand\h{\frac{1}{2}}

\renewcommand\l{\lambda}
\renewcommand\L{\Lambda}
\newcommand\m{\mu}
\newcommand\n{\nu}
\newcommand\om{\omega}

\newcommand\vp{\varphi}
\renewcommand\P{\Phi}
\newcommand\pa{\partial}

\newcommand\pr{\prime}

\newcommand\s{\sigma}

\newcommand\wti{\widetilde}

\newcommand\cA{{\mathcal A}}
\newcommand\cB{{\mathcal B}}

\newcommand\cR{{\mathcal R}}

\newcommand{\ct}[1]{\cite{#1}}
\newcommand{\bib}[1]{\bibitem{#1}}
\newcommand\PRD[3]{\textsl{Phys. Rev.} \textbf{D#1}, #3 (#2)}

\newcommand\CQG[3]{\textsl{Class. Quantum Grav.} \textbf{#1}, #3 (#2)}

\newcommand\IJMPA[3]{\textsl{Int. J. Mod. Phys.} \textbf{A#1}, #3 (#2)}

\newcommand\MPLA[3]{\textsl{Mod. Phys. Lett.} \textbf{A#1}, #3 (#2)}

\newcommand\vpdot{\stackrel{.}{\varphi}}
\newcommand\psidot{\stackrel{.}{\psi}}

\newcommand\adot{\stackrel{.}{a}}
\newcommand\addot{\stackrel{..}{a}}

\makeindex

\begin{document}

\sloppy \raggedbottom

\title{Quintessence in Multi-Measure Generalized Gravity Stabilized by 
Gauss-Bonnet/Inflaton Coupling}

%

\runningheads{Quintessence Stabilized by Gauss-Bonnet/Inflaton Coupling}{E. Guendelman, 
E. Nissimov and S. Pacheva}

\begin{start}

\coauthor{Eduardo Guendelman}{1,2,3}, \coauthor{Emil Nissimov}{4},
\coauthor{Svetlana Pacheva}{4}

\address{Department of Physics, Ben-Gurion Univ. of the Negev, \\
Beer-Sheva 84105, Israel}{1}

\address{Bahamas Advanced Study Institute and Conferences, 4A Ocean Heights, \\
Hill View Circle, Stella Maris, Long Island, The Bahamas}{2}

\address{Frankfurt Institute for Advanced Studies, Giersch Science Center, \\
Campus Riedberg, Frankfurt am Main, Germany}{3}

\address{Institute of Nuclear Research and Nuclear Energy, \\
Bulg. Acad. Sci., Sofia 1784, Bulgaria}{4}

\begin{Abstract}
We consider a non-standard generalized model of gravity coupled to a neutral scalar
``inflaton'' as well as to the fields of the electroweak bosonic sector.
The essential new ingredient is employing two alternative non-Riemannian 
space-time volume-forms (non-Riemannian volume elements, or covariant integration 
measure densitities) independent of 
the space-time metric. The latter are defined in terms of auxiliary
antisymmentric tensor gauge fields, which although not introducing any additional propagating
degrees of freedom, trigger a series of important features such as:
(i) appearance of two infinitely large flat regions of the effective ``inflaton'' 
potential in the corresponding Einstein frame with vastly different scales
corresponding to the ``early'' and ``late'' epochs of Universe's evolution;
(ii) dynamical generation of Higgs-like spontaneous symmetry breaking 
effective potential for the $SU(2)\times U(1)$ iso-doublet electroweak
scalar in the ``late'' universe, whereas it remains massless in the
``early'' universe.

Next, to stabilize the quintessential dynamics, we introduce in addition a
coupling of the ``inflaton'' to Gauss-Bonnet gravitational term. The latter
leads to the following radical change of the form of the total effective
``inflaton'' potential: its flat regions are now converted into a local
maximum corresponding to a ``hill-top'' inflation in the ``early'' universe
with no spontaneous breakdown of electroweak gauge symmetry
and, correspondigly, into a local minimum corresponding to the ``late''
universe evolution with a very small value of the dark energy and with
operating Higgs mechanism.
\end{Abstract}


\end{start}

\section{Introduction}
\label{intro}

The interplay between the cosmological dynamics and the evolution of the
symmetry breaking patterns along the history of the Universe is one of the
most important paradigms at the interface of particle physics and cosmology
\ct{general-cit}. Specifically, for the present epoch's phase of slowly accelerating 
Universe (dark energy domination) see \ct{dark-energy-observ} and for a recent 
general account, see \ct{rubakov-calcagni}.

Within this context, some of the main issues we will be addressing in the present
contribution are:

$\phantom{aaa}$(i) The existence of ``early'' Universe inflationary phase with 
unbroken electro-weak symmetry;

$\phantom{aaa}$(ii) The ``quintessential'' evolution towards ``late'' Universe 
epoch with a dynamically induced Higgs mechanism;

$\phantom{aaa}$(iii) Stability of the ``late'' Universe with spontaneous 
electro-weak breakdown and with a very small vacuum energy density via dynamically
generated cosmological constant.

Study of issues (i) and (ii) has already been initiated in 
Refs.\ct{grf-essay,BJP-3rd-congress}. Our approach is based on the powerful 
formalism of non-Riemannian volume-forms on the pertinent spacetime manifold
\ct{TMT-orig-1}-\ct{TMT-no-5th-force} (for further developments, 
see Refs.\ct{TMT-recent}). Non-Riemannian spacetime volume-forms
or, equivalently, alternative generally covariant integration measure densities
are defined in terms of auxiliary maximal-rank antisymmetric tensor gauge fields 
(``measure gauge fields'') unlike the standard Riemannian integration
measure density given given in terms of the square root of the determinant of the 
spacetime metric. These non-Riemannian-measure-modified gravity-matter models
are also called ``two-measure'', or more appropriately --
``multi-measure gravity theories''.  

The method of non-Riemannian spacetime volume-forms has profound impact in any 
(field theory) models with general coordinate reparametrization invariance, 
such as general relativity and its extensions
(\ct{TMT-orig-1}-\ct{buggy}), 
strings and (higher-dimensional) membranes \ct{mstring}, and supergravity
\ct{susyssb}, with the following  main features:

\begin{itemize}
\item 
Cosmological constant and other dimensionful constants are dynamically generated 
as arbitrary integration constants in the solution of the equations of motion for the 
auxiliary ``measure'' gauge fields. 
\item
An important characteristic feature is the global Weyl-scale 
invariance \ct{TMT-orig-2} of the starting Lagrangians actions of the underlying
generalized multi-measure gravity-matter models (for a similar recent
approach , see also \ct{hill-etal}). Global Weyl-scale symmetry is responsible for
the absence of a ``fifth force'' \ct{TMT-no-5th-force}. It undergoes
spontaneous breaking due to the appearance of the above mentioned
dynamically generated dimensionfull intergation constants.
\item
Applying the canonical Hamiltonian formalism for Dirac-constrained systems shows that 
the auxiliary ``measure'' gauge fields are in fact almost ``pure gauge'', which do not 
correspond to propagating field degrees of freedom. The only remnant of the
latter are the above mentioned arbitrary integration constants, which are identified 
with the conserved Dirac-constrained canonical momenta conjugated to certain components 
of the ``measure'' gauge fields \ct{quintess,buggy}.
\item
Applying the non-Riemannian volume-form formalism to minimal $N=1$ supergravity we 
arrive at a novel mechanism for the supersymmetric Brout-Englert-Higgs effect, namely, 
the appearance of a dynamically generated cosmological constant triggers spontaneous 
supersymmetry breaking and mass generation for the gravitino \ct{susyssb}. 
Applying the same non-Riemannian volume-form formalism to anti-de Sitter supergravity 
produces simultaneously a very large physical gravitino mass and a very small 
{\em positive} observable cosmological constant \ct{susyssb} in accordance 
with modern cosmological scenarios for slowly expanding universe of the present epoch
\ct{dark-energy-observ}. 
\item
Employing two different non-Riemannian volume-forms in generalized gravity-matter 
models thanks to the appearance of several arbitrary integration 
constants through the equations of motion w.r.t. the ``measure'' gauge fields, 
we obtain a remarkable effective scalar field potential with two infinitely large flat 
regions \ct{emergent,quintess} -- $(-)$ flat region for large negative values
and $(+)$ flat region for large positive values of the
scalar ``inflaton'' with vastly different energy scales -- appropriate for a
unified description of both the ``early'' and ``late'' Universe evolution.
An intriguing feature is the existence of a stable initial phase of
{\em non-singular} universe creation preceding the inflationary phase
-- stable ``emergent universe'' without ``Big-Bang'' \ct{emergent}.
\end{itemize}

In Section 2 below we describe the construction of a non-standard generalized model 
of gravity coupled to a neutral scalar ``inflaton'', as well as to the fields of the 
electroweak bosonic sector, employing the formalism of non-Riemannian
space-time volume forms. A crucial feature of the corresponding total
effective scalar field potential with the two infinitely large flat regions
is that in the $(-)$ flat region (``early'' Universe) the Higgs-like scalar
of the electro-weak sector remains massless (no Higgs mechanism), whereas in
the $(+)$ flat region (``late'' Universe) a Higgs-like effective potential is dynamically
generated triggering the standard electro-weak symmetry breaking.

A slightly different version of the formalism of Section 2 is briefly discussed 
in Appendix A -- it is inspired by Bekenstein's idea about gravity-assisted 
spontaneous electro-weak symmetry breakdown \ct{bekenstein-86}.

Next, in Section 3 we turn to the study of the stability issue (iii) formulated 
above. Namely, it is desirable that the ``late'' Universe epoch, instead of the
infinitely large $(+)$ flat region, would be described in terms of a stable
minimum of the effective ``inflaton'' potential. 
To this end we will introduce an additional {\em linear} coupling of the ``inflaton'' to 
Gauss-Bonnet gravitational term. 

The Gauss-Bonnet scalar density is a specific example of 
gravitational terms containing higher-order powers in the curvature
invariants, which appear naturally as renormalization counterterms in
quantized general relativity \ct{birrel-davis}, as well as in the context of string
theory \ct{GB-strings}.

Recently, within the standard Einstein general relativistic setting the role 
of Gauss-Bonnet-``inflaton'' couplings with various types
of functional dependence on the ``inflaton'' field has been extensively 
discussed in the cosmological context \ct{GB-cosmology}.

Previously, in \ct{E-N-S} some of us have studied a simplified generalized
gravity-scalar-field model based on a single non-Riemannian volume element with a
linear ``inflaton''-Gauss-Bonnet coupling. In the absence of Gauss-Bonnet
coupling the effective ``inflaton'' potential possesses in this case only
one infinitely long flat region with a very small height of the order of the
vacuum energy density in the ``late'' Universe. In the presence of the 
Gauss-Bonnet coupling, which modifies the ``inflaton'' effective potential,
one finds the appearance of a local minimum on top of the aforementioned 
flat region of the ``inflaton''  potential signalling stabilization of
the ``late'' Universe evolution with very small effective cosmological constant.

Here we will extend the work in \ct{E-N-S} by showing that the linear 
``inflaton''-Gauss-Bonnet coupling has a dramatic effect on 
the form of the total effective ``inflaton'' potential in the above mentioned
quintessence model based on generalized {\em multi-measure} gravity-matter theories
\ct{emergent,quintess} in the presence of the electro-weak bosonic sector 
\ct{grf-essay,BJP-3rd-congress}: 

$\phantom{aaa}$ (a) Its $(-)$ flat region is now converted into a local
maximum corresponding to a ``hill-top'' inflation in the ``early'' universe
a'la Hawking-Hertog mechanism \ct{hawking-hertog} with {\em no} spontaneous 
breakdown of electroweak gauge symmetry;

$\phantom{aaa}$ (b) Its $(+)$ flat region is converted into a local 
{\em stable minimum} corresponding to the ``late''
universe evolution with a very small value of the dark energy and with
operating standard Higgs mechanism.

In Appendix B we discuss a slightly different version of the formalism in
Section 3, where we will add a linear ``inflaton''-Gauss-Bonnet coupling
already to the initial action of the globally Weyl-scale invariant multi-measure
quintessence model -- this is unlike the formalism of Section 3, where the 
linear ``inflaton''-Gauss-Bonnet coupling term is added to the corresponding 
Einstein-frame action. Although within the formalism of Appendix B the linear 
``inflaton''-Gauss-Bonnet coupling does preserve the initial global Weyl-scale 
invariance, it exhibits a disadventage after the passage to the physical 
Einstein-frame since a combinantion involving one of the auxiliary ``measure'' 
gauge fields intended to remain ``pure gauge'' appears now as an additional 
propagating field-theoretic degree of freedom.

\section{Quintessence from Flat Regions of the Effective Inflaton Potential}
\label{flat-regions}

Let us consider, following \ct{emergent,grf-essay}, a multi-measure 
gravity-matter theory constructed in terms of two different non-Riemannian 
volume-forms (volume elements), where
gravity couples to a neutral scalar ``inflaton'' and the bosonic sector of
the standard electro-weak model (using units where $G_{\rm Newton} = 1/16\pi$):
\br
S = \int d^4 x\,\P (A) \Bigl\lb g^{\m\n} R_{\m\n}(\G) + L_1 (\vp,X) + 
L_2 (\s,X_\s;\vp) \Bigr\rb
\nonu \\
+ \int d^4 x\,\P (B) \Bigl\lb U(\vp) + L_3 (\cA,\cB) +
\frac{\P (H)}{\sqrt{-g}}\Bigr\rb \; .
\lab{TMMT-1}
\er
Here the following notations are used:
\begin{itemize}
\item
$\P(A)$ and $\P(B)$ are two independent non-Riemannian volume elements:
\be
\P (A) = \frac{1}{3!}\vareps^{\m\n\kappa\lambda} \pa_\m A_{\n\kappa\lambda} \quad ,\quad
\P (B) = \frac{1}{3!}\vareps^{\m\n\kappa\lambda} \pa_\m B_{\n\kappa\lambda} \; ,
\lab{Phi-1-2}
\ee
\item
$\P (H) = \frac{1}{3!}\vareps^{\m\n\kappa\lambda} \pa_\m H_{\n\kappa\lambda}$  
is the dual field-strength of an additional auxiliary tensor gauge field $H_{\n\kappa\lambda}$
crucial for the consistency of \rf{TMMT-1}.
\item
We are using Palatini formalism for the Einstein-Hilbert action: the scalar 
curvature is given by $R=g^{\m\n} R_{\m\n}(\G)$, where the metric $g_{\m\n}$
and the affine connection $\G^{\l}_{\m\n}$  are \textsl{a priori} independent.
\item
The ``inflaton'' Lagrangian terms are as follows:
\br
L_1 (\vp,X) = X - V_1 (\vp) \quad, \quad
X \equiv - \h g^{\m\n} \pa_\m \vp \pa_\n \vp \; ,
\lab{L-1} \\
V_1 (\vp) = f_1 \exp \{\a\vp\} \quad ,\quad U(\vp) = f_2 \exp \{2\a\vp\} \; ,
\lab{L-2}
\er
where $\a, f_1, f_2$ are dimensionful positive parameters.
\item
$\s \equiv (\s_a)$ is a complex $SU(2)\times U(1)$ iso-doublet scalar field
with the isospinor index $a=+,0$ indicating the corresponding $U(1)$ charge.
Its Lagrangian reads:
\be
L_2 (\s,X_\s;\vp) = X_\s - V_0 (\s) e^{\a\vp} \quad ,\quad 
X_\s \equiv - g^{\m\n} \bigl(\nabla_\m \s_a)^{*}\nabla_\n \s_a \; ,
\lab{L-sigma}
\ee
where the ``bare'' $\s$-field potential is of the same form as the standard Higgs
potential:
\be
V_0 (\s) = \frac{\l}{4} \((\s_a)^{*}\s_a - \m^2\)^2 \; .
\lab{standard-higgs}
\ee
In Appendix A below we will choose a different (simpler) version of 
$V_0 (\s)$ \rf{V0-bekenstein}.
\item
The gauge-covariant derivative acting on $\s$ reads:
\be
\nabla_\m \s = 
\Bigl(\pa_\m - \frac{i}{2} \tau_A \cA_\m^A - \frac{i}{2} \cB_\m \Bigr)\s \; ,
\lab{cov-der}
\ee
with $\h \tau_A$ ($\tau_A$ -- Pauli matrices, $A=1,2,3$) indicating the $SU(2)$ 
generators and $\cA_\m^A$ ($A=1,2,3$)
and $\cB_\m$ denoting the corresponding $SU(2)$ and $U(1)$ gauge fields.
\item
The gauge field kinetic terms in \rf{TMMT-1} are (all indices $A,B,C = (1,2,3)$):
\br
L_3 (\cA,\cB) = - \frac{1}{4g^2} F^2(\cA) - \frac{1}{4g^{\pr\,2}} F^2(\cB) \; ,
\lab{L-gaugefields} \\
F^2(\cA) \equiv F^A_{\m\n} (\cA) F^A_{\kappa\lambda} (\cA) g^{\m\kappa}
g^{\n\lambda} \; ,\;
F^2(\cB) \equiv F_{\m\n} (\cB) F_{\kappa\lambda} (\cB) g^{\m\kappa}
g^{\n\lambda} \; ,
\lab{F2-def} \\
F^A_{\m\n} (\cA) = 
\pa_\m \cA^A_\n - \pa_\n \cA^A_\m + \eps^{ABC} \cA^B_\m \cA^C_\n \; ,\;
F_{\m\n} (\cB) = \pa_\m \cB_\n - \pa_\n \cB_\m \; .
\lab{F-def}
\er
\end{itemize}

The form of the action \rf{TMMT-1} 
is fixed by the requirement of invariance under global Weyl-scale transformations:
\br
g_{\m\n} \to \lambda g_{\m\n} \;,\; \G^\m_{\n\lambda} \to \G^\m_{\n\lambda} \; ,\; 
\vp \to \vp - \frac{1}{\a}\ln \lambda\; 
\lab{scale-transf} \\
A_{\m\n\kappa} \to \lambda A_{\m\n\kappa} \; ,\; B_{\m\n\kappa} \to \lambda^2
B_{\m\n\kappa} \; ,\; 
H_{\m\n\kappa} \to H_{\m\n\kappa} \; ,
\nonu
\er
and the electro-weak fields remain inert under \rf{scale-transf}. 

Equations of motion for the affine connection $\G^\m_{\n\lambda}$ yield a solution for
the latter as a Levi-Civita connection:
\be
\G^\m_{\n\lambda} = \G^\m_{\n\lambda}({\bar g}) = 
\h {\bar g}^{\m\kappa}\(\pa_\n {\bar g}_{\lambda\kappa} + \pa_\lambda {\bar g}_{\n\kappa} 
- \pa_\kappa {\bar g}_{\n\lambda}\) \; ,
\lab{G-eq}
\ee
w.r.t. to the Weyl-conformally rescaled metric ${\bar g}_{\m\n}$:
\be
{\bar g}_{\m\n} = \chi_1 g_{\m\n} \quad ,\quad
\chi_1 \equiv \frac{\P_1 (A)}{\sqrt{-g}} \; . 
\lab{bar-g}
\ee
The metric ${\bar g}_{\m\n}$ plays an important role as the ``Einstein
frame'' metric (see \rf{einstein-frame} below).

Variation of the action \rf{TMMT-1} w.r.t. auxiliary tensor gauge fields
$A_{\m\n\lambda}$, $B_{\m\n\lambda}$ and $H_{\m\n\lambda}$ yields the equations:
\br
\pa_\m \Bigl\lb R + L_1 (\vp,X) + L_2 (\s,X_\s;\vp)\Bigr\rb = 0 \quad ,
\lab{A-eqs} \\
\pa_\m \Bigl\lb U(\vp) + L_3 (\cA,\cB) + \frac{\P (H)}{\sqrt{-g}}\Bigr\rb = 0 
\quad, \quad \pa_\m \Bigl(\frac{\P_2 (B)}{\sqrt{-g}}\Bigr) = 0 \; ,
\lab{B-H-eqs}
\er
whose solutions read:
\br
\frac{\P_2 (B)}{\sqrt{-g}} \equiv \chi_2 = {\rm const} \;\; ,\;\;
R + L_1 (\vp,X) + L_2 (\s,X_\s;\vp) = M_1 = {\rm const} \; ,
\nonu \\
U(\vp) + L_3 (\cA,\cB) +\frac{\P (H)}{\sqrt{-g}} = - M_2  = {\rm const} \; .
\lab{integr-const}
\er
Here $M_1$ and $M_2$ are arbitrary dimensionful and $\chi_2$
arbitrary dimensionless integration constants. We will take all $M_{1,2},\,\chi_2$
to be positive.

The first integration constant $\chi_2$ in \rf{integr-const} preserves
global Weyl-scale invariance \rf{scale-transf}
whereas the appearance of the second and third integration constants $M_1,\, M_2$
signifies {\em dynamical spontaneous breakdown} of global Weyl-scale invariance 
under \rf{scale-transf} 
due to the scale non-invariant solutions (second and third ones) in \rf{integr-const}. 

It is important to elucidate the physical meaning of the three arbitrary 
integration constants $M_1,\, M_2,\,\chi_2$ from the point of view of the
canonical Hamiltonian formalism. Namely, as shown in \ct{grf-essay}, 
the auxiliary maximal rank antisymmetric 
tensor gauge fields $A_{\m\n\lambda}, B_{\m\n\lambda}, H _{\m\n\lambda}$
entering the original non-Riemannian volume-form action \rf{TMMT-1} do 
{\em not} correspond to additional propagating field-theoretic degrees of freedom.
The integration constants $M_1,\, M_2,\,\chi_2$ are the only dynamical
remnant of the latter and they are identified  as conserved {\em Dirac-constrained}
canonical momenta conjugated to (certain components of) 
$A_{\m\n\lambda}, B_{\m\n\lambda}, H _{\m\n\lambda}$.

Following \ct{emergent,grf-essay} we first find from \rf{integr-const} 
the expression for $\chi_1$ \rf{bar-g} as algebraic function of the scalar matter
fields:
\be
\frac{1}{\chi_1} = \frac{1}{2\chi_2}\, 
\frac{V_1 (\vp) + V_0 (\s)e^{\a\vp} + M_1}{U(\vp) + M_2} \; .
\lab{chi1-eq}
\ee
Then we perform transition from the original metric $g_{\m\n}$ to ${\bar g}_{\m\n}$ 
arriving at the {\em ``Einstein-frame''} formulation, where the gravity equations 
of motion are written in the standard form of Einstein's equations:
\be
R_{\m\n}({\bar g}) - \h {\bar g}_{\m\n} R({\bar g}) = \h T^{\rm eff}_{\m\n} 
\lab{einstein-frame}
\ee
originating from the Einstein-frame action:
\be
S_{\rm EF} = \int d^4 x \sqrt{-{\bar g}} \Bigl\lb R({\bar g}) + 
L_{\rm eff}\bigl(\vp,{\bar X},\s,{\bar X}_\s,\cA,\cB\bigr)\Bigr\rb \; ,
\lab{einstein-frame-action}
\ee
with the {\em effective} energy-momentum tensor $T^{\rm eff}_{\m\n}$ given in terms
of the Einstein-frame matter Lagrangian $L_{\rm eff}$: 
\be
L_{\rm eff} = {\bar X} + {\bar X}_\s - U_{\rm eff}\bigl(\vp,\s\bigr)  
- \frac{\chi_2}{4g^2} {\bar F}^2(\cA) 
- \frac{\chi_2}{4g^{\pr\,2}} {\bar F}^2(\cB) \; .
\lab{L-eff}
\ee
Here bars indicate that the quantities are given in terms of the
Einstein-frame metric \rf{bar-g}, \textsl{e.g.}, 
${\bar X} \equiv - \h {\bar g}^\m\n \pa_\m \vp \pa_\n \vp$ , 
${\bar X}_\s \equiv - {\bar g}^{\m\n} \bigl(\nabla_\m \s_a)^{*}\nabla_\n\s_a$ 
{\em etc}, and the total scalar field effective potential reads:
\br
U_{\rm eff}\bigl(\vp,\s\bigr) = 
\frac{\Bigl(V_1 (\vp) + V_0 (\s)e^{\a\vp} + M_1\Bigr)^2}{4\chi_2 (U(\vp) + M_2)}
\nonu \\
= \frac{\Bigl( M_1 e^{-\a\vp} + f_1 + \frac{\lambda}{4} \((\s_a)^{*}\s_a - \m^2\)^2
\Bigr)^2}{4\chi_2 (M_2 e^{-2\a\vp} + f_2)}
\lab{U-eff-total}
\er
(see Eq.\rf{U-eff-total-0} below for the Bekenstein-inspired form of $V_0 (\s)$
\rf{V0-bekenstein}).

A remarkable feature of the effective scalar potential $U_{\rm eff} (\vp,\s)$ 
\rf{U-eff-total} is that it possesses two {\em infinitely large flat regions}
describing the ``early'' and ``late'' Universe, respectively (see 
\rf{U-plus-magnitude} and \rf{U-minus-magnitude} below):

\begin{itemize}
\item
{\em (-) flat region} -- for large negative values of $\vp$, where: 
\be
U_{\rm eff}(\vp,\s) \simeq U_{(-)} \equiv \frac{M_1^2}{4\chi_2\,M_2} \; .
\lab{U-minus} 
\ee
In this region the Higgs-like field $\s$ remains massless and there is {\em no
spontaneous breakdown} of electro-weak gauge symmetry.
\item
{\em (+) flat region} -- for large positive values of $\vp$, where:
\be
U_{\rm eff}(\vp,\s) \simeq U_{(+)}(\s) = 
\frac{\Bigl(\frac{\lambda}{4} \((\s_a)^{*}\s_a - \m^2\)^2 + f_1\Bigr)^2}{4\chi_2\,f_2} \; ,
\lab{U-plus}
\ee
which obviously yields as a lowest lying vacuum the Higgs one $|\s| = \m$
with a residual effective cosmological constant $\L_{(+)}$:
\be
2\L_{(+)} \equiv U_{(+)}(\m) = \frac{f_1^2}{4\chi_2 f_2} \; .
\lab{CC-eff-plus}
\ee
For the Bekenstein-inspired form of $U_{(+)}(\s)$, see Eq.\rf{U-plus-0} below.
\end{itemize}

Choosing the scales of the original ``inflaton'' coupling constants $f_{1,2}$ in
terms of fundamental physical constants as:
\be
f_1 \sim M^4_{EW} \quad , \quad f_2 \sim M^4_{Pl} \; ,
\lab{f12-scales}
\ee
where $M_{EW},\, M_{Pl}$ are the electroweak and Plank scales,
respectively, we are then naturally led to a very small vacuum energy
density in the {\em (+) flat region} \rf{CC-eff-plus}:
\be
U_{(+)}(\m) \sim M^8_{EW}/M^4_{Pl} \sim 10^{-122} M^4_{Pl} \; ,
\lab{U-plus-magnitude}
\ee
which is the right order of magnitude for the present epoch's (``late'' Universe) 
vacuum energy density as already realized in \ct{arkani-hamed}.

On the other hand, if we take the order of magnitude of the integration
constants 
\be
M_1 \sim M_2 \sim 10^{-8} M_{Pl}^4 \; ,
\lab{M12-scales}
\ee
then the order of magnitude 
of the vacuum energy density in the {\em (-) flat region} \rf{U-minus} becomes:
\be
U_{(-)} \sim M_1^2/M_2 \sim 10^{-8} M_{Pl}^4 \; ,
\lab{U-minus-magnitude}
\ee
which conforms to the Planck Collaboration data \ct{Planck} for the
``early'' Universe's energy scale of inflation being of order $10^{-2} M_{Pl}$.

\begin{figure}
\begin{center}
\includegraphics{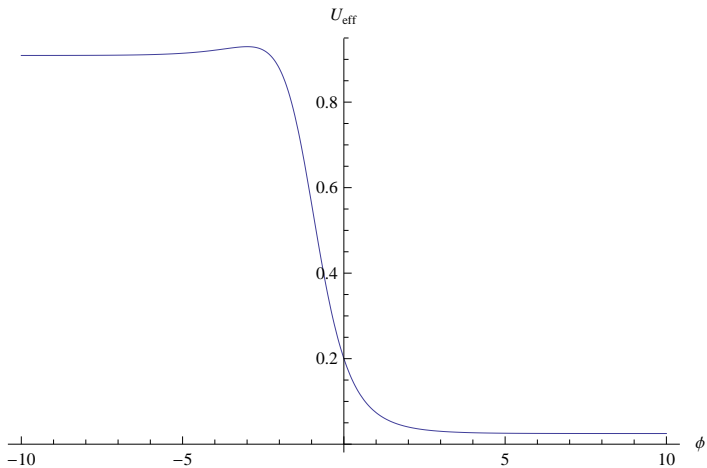}
\caption{Qualitative shape of the effective ``inflaton'' potential $U_{\rm eff}$
\rf{U-eff-total} as function of $\vp$ (for fixed $\s$) before inflaton coupling to 
Gauss-Bonnet term.}
\end{center}
\end{figure}

Let us note the small ``bump'' on the l.h.s. of the graph (Fig.1) of $U_{\rm eff}$
\rf{U-eff-total} as function of $\vp$ and where $|\s_{\rm vac}|=0$ -- this is a local 
maximum located towards the end of the $(-)$ flat region at $\vp = \vp_{\rm max}$:
\be
e^{-\a \vp_{\rm max}}= \frac{M_1 f_2}{M_2 f_1(\m)} \quad, \quad
f_1(\m) \equiv f_1 +\lambda\m^4/4 \; .
\lab{bump}
\ee
We note that the relative height $\Delta U_{(-)}$ of the above mentioned 
``bump'' of the inflaton potential \rf{U-eff-total} (at $|\s_{\rm vac}|=0$) 
w.r.t. the height of the $(-)$ flat region \rf{U-minus}:
\be
\Delta U_{(-)} \equiv U_{\rm eff} (\vp_{\rm max},0) - \frac{M_1^2}{4\chi_2 M_2}
= \frac{f_1^2 (\m)}{4\chi_2 f_2}
\lab{delta-bump}
\ee
is of the same order of magnitude as the small effective cosmological constant 
\rf{CC-eff-plus} in the $(+)$ flat region (``late'' Universe) (recall 
$f_1 \sim M_{EW}$, $\m \sim M_{EW}$ and the bare Higgs-like dimensionless self-coupling 
$\lambda$ being small).

On the other hand, the inflaton potential \rf{U-eff-total} at 
$|\s|= |\s_{\rm vac}| = \m$ does not possess a strict minimum on the $(+)$ flat 
region -- the strict minimum occurs formally at $\vp \to +\infty$. 
In the next Section we will see how adding a
coupling of the inflaton to a gravitational Gauss-Bonnet density will 
convert the infinitely large $(+)$ flat region of the effective inflaton
potential into a region with a stable minimum. Simultaneously, the 
infinitely large $(-)$ flat region of the effective inflaton potential with
the small ``bump'' at its end \rf{bump}-\rf{delta-bump} will be converted into 
a region with well-peaked maximum and sharper decent for large negative inflaton values
(see Fig.3 below).

\section{Adding Gauss-Bonnet/Inflaton Coupling}
\label{GB-coupling}

Let us now supplement the Einstein-frame action \rf{einstein-frame-action} with a 
linear coupling of the ``inflaton'' to gravitational Gauss-Bonnet term
$\cR_{\rm GB}$ with a (positive) coupling constant $b$:
\br
S_{\rm EF} = \int d^4 x \sqrt{-{\bar g}} \Bigl\lb R({\bar g}) + 
{\bar X} + {\bar X}_\s - U_{\rm eff}\bigl(\vp,\s\bigr) 
\nonu \\
- \frac{\chi_2}{4g^2} {\bar F}^2(\cA) 
- \frac{\chi_2}{4g^{\pr\,2}} {\bar F}^2(\cB)
 - b\, \vp\, {\bar\cR}_{\rm GB} \Bigr\rb\; ,
\lab{einstein-frame-GB}
\er
with $U_{\rm eff}\bigl(\vp,\s\bigr)$ as in \rf{U-eff-total}, and:
\be
{\bar \cR}_{\rm GB} = {\bar R}_{\m\n\kappa\lambda} {\bar R}^{\m\n\kappa\lambda} 
- 4 {\bar R}_{\m\n} {\bar R}^{\m\n} + {\bar R}^2 \; ,
\lab{GB-def}
\ee
where all objects with superimposed bars are defined w.r.t. second-order formalism 
with the Einstein-frame metric ${\bar g}_{\m\n}$.

Here we will be interested in ``vacuum'' solutions, \textsl{i.e.}, for
constant values of the matter fields. The corresponding equations of motion for 
constant $\vp$ and $\s$ read:
\be
{\bar R}_{\m\n} - \h {\bar g}_{\m\n} {\bar R} = 
-\h {\bar g}_{\m\n} U_{\rm eff}(\vp,\s) \; ,
\lab{g-eqs}
\ee
note that the Gauss-Bonnet coupling does {\em not} contribute to the vacuum
energy density on the r.h.s. of \rf{g-eqs};
\br
\partder{}{\vp} U_{\rm eff}(\vp,\s) + b \cR_{\rm GB} = 0 \; ;
\lab{vp-eq} \\
\partder{}{\s_a} U_{\rm eff}(\vp,\s) = 0 \quad \longrightarrow \quad
\partder{}{\s_a} V_0 (\s) = 0 \; \phantom{aaa}
\nonu \\
\longrightarrow \quad (\s_a)^{*} \((\s_{a^\pr})^{*}\s_{a^\pr} - \m^2\) = 0 \;\;
\longrightarrow \;\;\;|\s_{\rm vac}|=\m \;\; {\mathrm or}\;\; |\s_{\rm vac}|=0 \; .
\lab{sigma-eq}
\er
For constant $\vp$ and $\s$ the solution to \rf{g-eqs} is maximally symmetric:
\be
{\bar R}_{\m\n\kappa\lambda} = \frac{1}{6} U_{\rm eff}(\vp,\s) 
\bigl({\bar g}_{\m\kappa} {\bar g}_{\n\lambda} - {\bar g}_{\m\lambda} {\bar g}_{\n\kappa}\bigr)
\; ,
\lab{max-symm}
\ee
which yields for the Gauss-Bonnet term \rf{GB-def}:
\be
\cR_{\rm GB} = \frac{2}{3} \Bigl( U_{\rm eff}(\vp,\s)\Bigr)^2 \; .
\lab{GB-vac}
\ee
Inserting \rf{GB-vac} into  $\vp$-``vacuum'' equation \rf{vp-eq} we get:
\be
\partder{}{\vp} U_{\rm eff}(\vp,\s_{\rm vac}) 
+ \frac{2b}{3}\Bigl( U_{\rm eff}(\vp,\s_{\rm vac})\Bigr)^2 = 0 \; ,
\lab{vp-vac-eq}
\ee
with $\s_{\rm vac}$ as in \rf{sigma-eq}. Eq.\rf{vp-vac-eq} implies that in
fact the total effective inflaton potential after introducing
Gauss-Bonnet/inflaton linear coupling is modified from 
$U_{\rm eff}(\vp,\s_{\rm vac})$ \rf{U-eff-total} to the following one:
\be
V_{\rm total}(\vp,\s_{\rm vac}) = U_{\rm eff}(\vp,\s_{\rm vac})
+ \frac{2b}{3}\int^{\vp} d\phi \Bigl( U_{\rm eff}(\phi,\s_{\rm vac})\Bigr)^2 \; ,
\lab{V-total-GB}
\ee

Eq.\rf{vp-vac-eq} upon inserting the explicit expression \rf{U-eff-total}
acquires the form:
\br
\partder{}{\vp} V_{\rm total}(\vp,\s_{\rm vac})=
\frac{b\, M_1^4 \bigl(e^{-\a\vp} + {\wti f}_1/M_1\Bigr)}{24 \chi_2^2 M_2^2
\bigl(e^{-2\a\vp}+ f_2/M_2\Bigr)}\, F(e^{-\a\vp}) = 0 \; ,
\lab{vp-extrema} \\
{\wti f}_1 \equiv f_1 + \frac{\lambda}{4} (|\s_{\rm vac}|^2 - \m^2)^2 =
\left\{\begin{array}{ll} {f_1 \quad {\mathrm for}\;\; |\s_{\rm vac}|=\m} \\ 
{f_(\m) \equiv f_1 + \frac{\lambda}{4}\m^4} \quad {\mathrm for}\;\; \s_{\rm vac} = 0
\end{array} \right.
\lab{f1-tilde}
\er
where the ``vacuum'' solutions $z \equiv e^{-\a\vp_{\rm vac}}$ must be real
positive roots of the following cubic polynomial:
\be
F(z) \equiv z^3 + 
\frac{3{\wti f}_1}{M_1}\Bigl( 1 + \frac{4\a\chi_2 M_2}{b\, M_1^2}\Bigr) z^2
- \frac{3{\wti f}^2_1}{M_1^2}\Bigl(\frac{4\a\chi_2 f_2}{b\,{\wti f}^2_1} - 1\Bigr) z
+ \frac{{\wti f}^3_1}{M_1^3} = 0 \; .
\lab{F-def-eq}
\ee

Existence of two {\em different} positive roots of $F(z)$ \rf{F-def-eq} -- 
$z_0 (b) \equiv e^{-\a\vp_0 (b)}$ corresponding to a minimum of
$V_{\rm total}(\vp,\s_{\rm vac})$
\rf{V-total-GB}, and $z_1 (b) \equiv e^{-\a\vp_1 (b)}$ corresponding to a maximum of
$V_{\rm total}(\vp,\s_{\rm vac})$, where the dependence on the inflaton-Gauss-Bonnet 
coupling constant $b$ is explicitly indicated (\textsl{cf.} Fig.2 and 
Eqs.\rf{second-der-vp}-\rf{min-max} below)  -- imposes
the following upper limit for the parametric dependence of $\vp_{0,1}(b)$ on $b$:
\be
b < b_{\rm max} \equiv 
\frac{12\a\chi_2 M_2\,Q}{M_1^2 \lb 2Q^3 + 3Q^2 -3Q-2+2(Q^2 +Q+1)^{3/2}\rb}
\quad , \;\; Q\equiv \frac{M_2 {\wti f}^2_1}{M_1^2 f_2} \; .
\lab{b-limit}
\ee

\begin{figure}
\begin{center}
\includegraphics{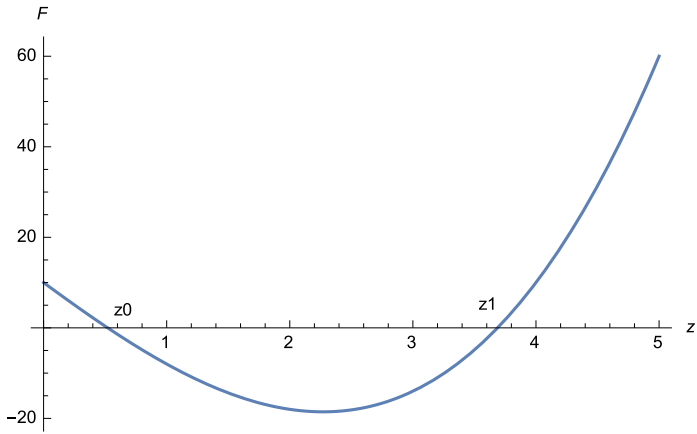}
\caption{Qualitative plot of the cubic polynomial $F(z)$ \rf{F-def-eq}.}
\end{center}
\end{figure}

The extremums $z_{0,1}(b) \equiv \exp\{-\a\,\vp_{0,1}(b)\}$ of 
$V_{\rm total}(\vp,\s_{\rm vac})$ \rf{V-total-GB} are given explicitly 
(for $0\leq b < b_{\rm max}$ \rf{b-limit}) as:
\br
z_{0,1}(b) = \sqrt{A} \Bigl\lb \cos\bigl(\frac{1}{3}\arctan\sqrt{A^3/B^2 - 1}\bigr)
\mp \sqrt{3} \sin\bigl(\frac{1}{3}\arctan\sqrt{A^3/B^2 - 1}\bigr)\Bigr\rb
\nonu \\
- \frac{{\wti f}_1}{M_1}\Bigl( 1 + \frac{4\a\chi_2 M_2}{b\, M_1^2}\Bigr) 
\quad , \phantom{aaaaa} 
\lab{z-0-1}
\er
where the quantities $A$ and $B$ are expressed in terms of the parameters as:
\br
A\equiv \frac{{\wti f}^2_1}{M^2_1}\,\om\,\bigl(1+2Q+Q^2\om\bigr) \quad ,
\quad \om \equiv \frac{4\a\chi_2 f_2}{b\, {\wti f}^2_1} \;\; ,\;\; 
Q \; {\mathrm as ~in ~\rf{b-limit}} \; ,
\lab{A-def}\\
B\equiv \frac{{\wti f}^3_1}{M^3_1}\,\om\,
\bigl\lb \frac{3}{2} + \frac{3}{2}Q(\om +1) + 3Q^2 \om + Q^3\om^2\bigr\rb \; .
\lab{B-def}
\er

The condition \rf{b-limit} comes from the inequality $A^3/B^2 - 1>0$ in \rf{z-0-1}.

For $b > b_{\rm max}$ there are no real positive roots of $F(z)$
\rf{F-def-eq}, and in the limiting case $b=b_{\rm max}$ the roots 
$z_{0,1}(b_{\rm max}) \equiv \exp\{-\a\vp_{0,1}(b_{\rm max})$ coalesce
and become an inflex point of $F(z)$ \rf{F-def-eq}:
\br
z_0 (b_{\rm max})= z_1 (b_{\rm max}) \equiv z (b_{\rm max})  
\quad ,\quad F^\pr \bigl( z(b_{\rm max})\bigr) = 0 \; ,
\lab{inflex} \\
z (b_{\rm max}) = 
\frac{{\wti f}_1}{M_1} \Bigl\lb \sqrt{(1+ Q\om_{\rm max})^2 + \om_{\rm max} -1}
- (1+ Q\om_{\rm max})\Bigr\rb \;\;,
\lab{z-max} \\
\om_{\rm max} \equiv \frac{4\a\chi_2 f_2}{b_{\rm max}\, {\wti f}^2_1} \; ,
\nonu
\er
using the short-hand notations in \rf{b-limit}, \rf{A-def}.
In other words, for $b \geq b_{\rm max}$ there are {\rm no}
extremums of the total inflaton effective potential \rf{V-total-GB}.

The second derivative w.r.t. $\vp$ of $V_{\rm total}(\vp,\s_{\rm vac})$ 
\rf{V-total-GB} at the extremums $z_{0,1}(b) \equiv \exp\{-\a\,\vp_{0,1}(b)\}$ reads:
\br
\frac{\pa^2}{\pa \vp^2} V_{\rm total}(\vp_{0,1},\s_{\rm vac}) =
- \frac{b\,\a z_{0,1}(b) M_1^4 \bigl( z_{0,1}(b)+{\wti f}_1/M_1\bigr)}{
24\chi_2^2 M_2^2 \bigl( z^2_{0,1}(b)+ f_2/M_2\bigr)^2}\, F^{\pr}\bigl(z_{0,1}(b)\bigr) \; ,
\lab{second-der-vp} \\
F^{\pr}\bigl(z_{0,1}(b)\bigr) = 3 z^2_{0,1}(b) + 
\frac{6{\wti f}_1}{M_1}\Bigl( 1 + \frac{4\a\chi_2 M_2}{b\, M_1^2}\Bigr) z_{0,1}(b)
\phantom{aaaaaaa}
\nonu \\
- \frac{3{\wti f}^2_1}{M_1^2}\Bigl(\frac{4\a\chi_2 f_2}{b\,{\wti f}^2_1} - 1\Bigr)
\; , \phantom{aaaaa}
\lab{F-der}
\er
where we have (see Fig.2):
\be
F^{\pr}\bigl(z_0 (b)\bigr) < 0 \quad ,\quad F^{\pr}\bigl(z_1(b)\bigr) > 0 \; .
\lab{min-max}
\ee
Taking also into account that:
\be
\frac{\pa^2}{\pa \s^2} U_{\rm eff}(\vp_0 (b),\s_{\rm vac}=\m) > 0 \quad ,\quad
\frac{\pa^2}{\pa \s^2} U_{\rm eff}(\vp_1 (b),\s_{\rm vac}=0) < 0 \; ,
\lab{second-der-sigma}
\ee
we conclude that (see Fig.3):

\begin{figure}
\begin{center}
\includegraphics{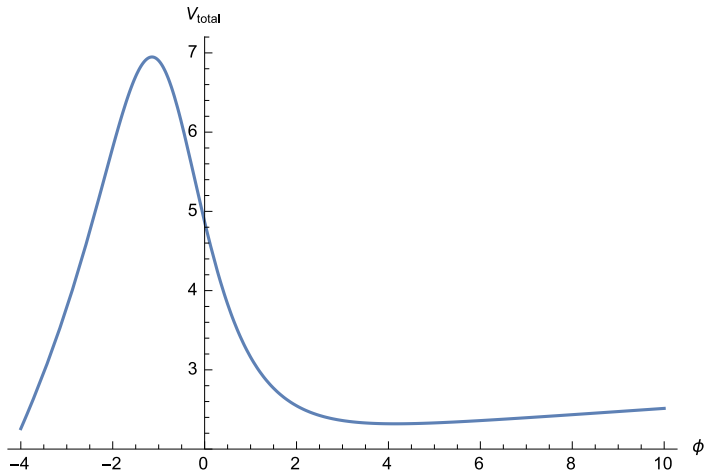}
\caption{Qualitative shape of the total effective ``inflaton'' potential
$V_{\rm total}(\vp,\s_{\rm vac})$ \rf{V-total-GB} as function of $\vp$ after
adding inflaton coupling to Gauss-Bonnet term.}
\end{center}
\end{figure}

\begin{itemize}
\item
$z_0 (b) \equiv e^{-\a\vp_0 (b)}$ \rf{z-0-1} with $\s_{\rm vac}=\m$ (spontaneous
breakdown of electro-weak symmetry) is a local {\em stable minimum} of the total
inflaton effective potential \rf{V-total-GB}. With the choice from Section 2
($f_1 \sim M^4_{EW}$, $f_2 \sim M^4_{Pl}$, $M_{1,2} \sim 10^{-8} M^4_{Pl}$)
we find ($b_{\rm max}$ as in \rf{b-limit}, $z (b_{\rm max})$ as in \rf{z-max}):
\be
0 \leq z_0 (b) \equiv e^{-\a\vp_0 (b)} < z (b_{\rm max}) \; ,
\lab{z0-inequal}
\ee
 where:
\be
z_0 (b) \equiv e^{-\a\vp_0 (b)}  \to 0 \;\;, {\mathrm i.e.} \;\; \vp_0 (b) \to +\infty 
\quad {\mathrm for}\; b \to 0 \; ,
\lab{z0-limit}
\ee
\textsl{i.e.}, recovering the $(+)$ flat region -- r.h.s. of Fig.1, and:
\be
\vp_0 (b) \to \frac{1}{\a}\, \log (z^{-1}(b_{\rm max}) 
\;\;\; {\mathrm for} \; b \to b_{\rm max}\; .
\lab{vp0-limit}
\ee
\item
$z_1 (b) \equiv e^{-\a\vp_1 (b)}$ \rf{z-0-1} with $\s_{\rm vac}=0$ ({\em no} 
spontaneous breakdown of electro-weak symmetry) is a local {\em maximum} of the total
inflaton effective potential \rf{V-total-GB}. Also we find here
($\vp_{\rm max}$ as in \rf{bump}): 
\be
z (b_{\rm max}) < z_1 (b) \equiv e^{-\a\vp_1 (b)} 
\leq e^{-\a\vp_{\rm max}} \equiv \frac{M_1 f_2}{M_2 f_1 (\m)} \; 
\lab{z1-inequal}
\ee
where:
\be
\vp_1 (b) \to \vp_{\rm max} \equiv 
- \frac{1}{\a}\log \frac{M_1 f_2}{M_2 f_1 (\m)}
\quad {\mathrm for}\; b \to 0 \; ,
\lab{z1-limit}
\ee
\textsl{i.e.}, recovering the $(-)$ flat region -- l.h.s. of Fig.1, and:
\be
\vp_1 (b) \to \frac{1}{\a}\, \log (z^{-1}(b_{\rm max})
\;\;\; {\mathrm for} \; b \to b_{\rm max}\; .
\lab{vp1-limit}
\ee
\end{itemize}

Let us also note the linear asymptotic behaviour of the total effective inflaton
potential \rf{U-eff-total} for very large positive and negative values of
the inflaton as it follows from \rf{V-total-GB} and 
\rf{CC-eff-plus}-\rf{U-minus-magnitude}:
\br
V_{\rm total}(\vp,\m) \to U_{(+)}(\m) 
+ \vp\,\frac{2b}{3} \bigl( U_{(+)}(\m)\bigr)^2 \quad, \;\;
{\mathrm for} \; \vp \to +\infty \; ,
\lab{asymptot-plus} \\
V_{\rm total}(\vp,0) \to U_{(-)} - |\vp|\,\frac{2b}{3} \bigl(U_{(-)}\bigr)^2
\quad,\;\; {\mathrm for} \; \vp \to -\infty \; .
\lab{asymptot-minus}
\er

\section{Discussion}
\label{discuss}

According to Eqs.\rf{g-eqs}, \rf{U-eff-total} the vacuum energy density 
at the stable minimum of the total inflaton effective potential \rf{V-total-GB} at
$z_0 (b) \equiv e^{-\a\vp_0 (b)}$ \rf{z-0-1}:
\be
U_{\rm eff}(\vp_0 (b),\m) = 
\frac{\Bigl(f_1(\m) + M_1 z_0 (b)\Bigr)^2}{4\chi_2\bigl(f_2 + M_2 z^2_0(b)\bigr)}
\lab{density-min}
\ee
is, according to \rf{z0-inequal}-\rf{vp0-limit} and \rf{z-max}, of the same order of
magnitude as the height \rf{CC-eff-plus} (vacuum energy density) of the $(+)$ 
flat region of the inflaton potential in the absence of Gauss-Bonnet coupling,
\textsl{i.e.}, it matches the vacuum energy density of the ``late'' Universe.
Now, however, due to the inflaton-Gauss-Bonnet coupling we have a small effective 
inflaton mass-squared $V^{\pr\pr}_{\rm total}(\vp_0 (b),\m)$ 
\rf{second-der-vp}-\rf{min-max}
(taking into account the orders of magnitude of $f_{1,2}$ and $M_{1,2}$).

According to the ``hill-top'' mechanism of Hawking-Hertog \ct{hawking-hertog}, 
the maximum of the total effective inflaton potential \rf{V-total-GB}
at $z_1 (b) \equiv e^{-\a\vp_1(b)}$ \rf{z-0-1} can be associated with the start
of inflation in the ``early'' Universe. One prerequisite of the latter is
smoothness of the maximum, \textsl{i.e.}, $-V^{\pr\pr}_{\rm total}(\vp_1 (b),0)$ 
\rf{second-der-vp}-\rf{min-max} should be small. The latter condition is
consistent only for small inflaton-Gauss-Bonnet coupling $b \ll b_{max}$,
since the vacuum energy density at the maximum
$z_1 (b) \equiv e^{-\a\vp_1(b)}$ \rf{z-0-1}:
\be
U_{\rm eff}(\vp_1 (b),0) = 
\frac{\Bigl(f_1 + M_1 z_1 (b)\Bigr)^2}{4\chi_2\bigl(f_2 + M_2 z^2_1(b)\bigr)}
\lab{density-max}
\ee
sharply diminishes from $U_{(-)}$ \rf{U-minus} at $b=0$ with $b$ growing towards 
$b_{max}$ and at 
$b\!simeq\! b_{\rm max}$, due to a coalescence of the minimum and the
maximum $z_0 (b_{\rm max})= z_1 (b_{\rm max})$ \rf{inflex}-\rf{z-max},
$U_{\rm eff}(\vp_1 (b_{\rm max}),0)$ becomes of the same order of magnitude as 
the vacuum energy density $U_{\rm eff}(\vp_0 (b_{\rm max}),\m)$ in the 
``late'' Universe.

The next task will be analyzing the corresponding Friedman equations upon
FLRW (Friedman-Lemaitre-Robertson-Walker) reduction of the Einstein-frame
metric (${\bar g}_{\m\n}dx^\m dx^\n = -dt^2 + a(t) d{\vec x}^2$, 
$H\equiv \frac{\adot}{a}$; recall Newton constant $G_N = 1/16\pi$).
Ignoring for simplicity the electro-weak gauge bosons, the Friedman
equations read:
\br
H^2 = \frac{1}{6} \rho_{\rm eff} \quad ,\quad
\rho_{\rm eff} \equiv \rho + 24 b\,\vpdot H^3 \; ,
\lab{friedman-1} \\
12 \frac{\addot}{a} + 3 p_{\rm eff} + \rho_{\rm eff} = 0 \; ,
\lab{friedman-2} 
\er
with:
\br
p_{\rm eff} \equiv \bigl( 1 - 4b\vpdot H +48 b^2 H^4\bigr)^{-1} 
\bigl\lb p + 32 b \vpdot H^3 + 8 b H^2 \frac{\pa U_{\rm eff}} {\pa\vp}
- 96 b^2 H^6 \bigr\rb \; ,
\nonu \\
\phantom{aaaa}
\lab{p-eff-def}
\er
where the second Friedman Eq.\rf{friedman-2} can be equivalently written as:
\be
4 \frac{d}{dt}\bigl( H - 2b\vpdot H^2\bigr) + \rho + p + 8b \vpdot H^3 = 0
\lab{friedman-2a} \; ,
\ee
and the ``inflaton'' equation of motion being:
\be
\frac{d}{dt}\bigl(\vpdot + 8b H^3\bigr) + 3H\bigl(\vpdot + 8b H^3\bigr)
+ \frac{\pa U_{\rm eff}}{\pa\vp} = 0 \; ,
\lab{vp-eq-FRLW}
\ee
where $U_{\rm eff}$ is as in \rf{U-eff-total} and $\rho$ and $p$ are the ordinary 
Einstein-frame matter energy density and pressure in the absence of 
inflaton-Gauss-Bonnet coupling.

\section*{Acknowledgements}
We gratefully acknowledge support of our collaboration through the academic exchange 
agreement between the Ben-Gurion University and the Bulgarian Academy of Sciences.
E.N. has received partial support from European COST actions MP-1405 and CA-16104. 
S.P. and E.N. are also partially supported by a Bulgarian National Science Fund 
Grant DFNI-T02/6. E.G. acknowledges partial support from European COST actions
CA-15117 and CA-16104. He is also grateful to Foundational Questions Institute (FQXi) 
for financial help through a FQXi mini grant to the Bahamas Advanced Study 
Institute's Conference 2017.

\appendix
\section*{Appendix A}

In a previous papers of ours \ct{grf-essay} we implemented an intriguing
idea of Bekenstein \ct{bekenstein-86} about a gravity-assisted spontaneous symmetry 
breaking of electro-weak (Higgs) type without invoking unnatural (according to
Bekenstein's opinion) ingredients like negative mass squared and a quartic
self-interaction for the Higgs field.

Instead of \rf{TMMT-1} (which appeared later in \ct{BJP-3rd-congress}) we first 
proposed in \ct{grf-essay} the following generalized gravity-matter action
(with some minor updates in the notations from \rf{TMMT-1}):
\br
{\widehat S} = \int d^4 x\,\P (A) \Bigl\lb g^{\m\n} R_{\m\n}(\G) 
- 2\L_0 \frac{\P (A)}{\sqrt{-g}} + X + {\widehat f}_1 e^{\a\vp}
+ {\widehat L}_2 (\s,X_\s;\vp) \Bigr\rb
\nonu \\
+ \int d^4 x\,\P (B) \Bigl\lb U(\vp) + L_3 (\cA,\cB) +
\frac{\P (H)}{\sqrt{-g}}\Bigr\rb \; . \phantom{aaaa}
\lab{TMMT-0}
\er
where:
\begin{itemize}
\item
Instead of $L_2 (\s,X_\s;\vp)$ \rf{L-sigma} we have in \rf{TMMT-0}:
\br
{\widehat L}_2 (\s,X_\s;\vp) = X_\s - V_0 (\s) e^{\a\vp} \quad ,\quad 
X_\s \equiv - g^{\m\n} \bigl(\nabla_\m \s_a)^{*}\nabla_\n \s_a \;\; ,
\lab{Lhat-sigma} \\
V_0 (\s) = m_0^2\,(\s_a)^{*}(\s_a) \;\; , \phantom{aaaa}
\lab{V0-bekenstein}
\er
where $V_0 (\s)$ is the standard (bare) mass term for the Higgs-like field 
$\s$ and there is no (bare) quartic self-interaction for the latter.
\item
Here we have an additional term qudratic w.r.t. the first non-Riemannian
volume-form density $\P (A)$ \rf{Phi-1-2} with a very small parameter 
$\L_0$ later to be identified with the present (``late'' Universe) epoch small 
observable cosmological constant.
\end{itemize}

Following the same steps as in Section 2 we obtain Einstein-frame effective
action of the same form as \rf{einstein-frame-action}-\rf{L-eff}, where now
the effective scalar potential reads:
\be
U_{\rm eff}\bigl(\vp,\s\bigr) = \frac{\Bigl( M_1 e^{-\a\vp}
- {\widehat f}_1 + m_0^2\,(\s_a)^{*}(\s_a)\Bigr)^2}{4\chi_2 (M_2 e^{-2\a\vp} + f_2)}
+ 2 \L_0 \; .
\lab{U-eff-total-0}
\ee
It obviously possesses again two infinitely large flat regions, where:
\begin{itemize}
\item
The $(-)$ flat region (for large negative values of $\vp$) of \rf{U-eff-total-0}
is almost the same  as in \rf{U-minus}:
\be
U_{\rm eff}(\vp,\s) \simeq U_{(-)} \equiv \frac{M_1^2}{4\chi_2\,M_2} + 2\L_0\; ,
\lab{U-minus-0} 
\ee
therefore here the Higgs-like field $\s$ remains massless.
\item
In the $(+)$ flat region (for large positive values of $\vp$) of
\rf{U-eff-total-0}:
\be
U_{\rm eff}(\vp,\s) \simeq U_{(+)}(\s) = 
\frac{\Bigl(-{\widehat f}_1 + m_0^2\,(\s_a)^{*}(\s_a)\Bigr)^2}{4\chi_2\,f_2} 
+ 2 \L_0 \; .
\lab{U-plus-0}
\ee
Spontaneous electro-weak symmetry breaking occurs at the ``vacuum'' value:
\be
|\s_{\rm vac}|=\frac{1}{m_0}\sqrt{{\widehat f}_1} \; ,
\lab{sigma-vac}
\ee
where the parameters are naturally identified as:
\be
{\widehat f}_1 \sim M^4_{EW} \quad ,\quad m_0 \sim M_{EW}
\lab{f1-m0-scales}
\ee
in terms of the electro-weak energy scale. Thus, the residual cosmological
constant $\L_0$ has to be identified with the current epoch observable 
cosmological constant ($\sim 10^{-122} M^4_{Pl}$) and, therefore, according 
to \rf{U-minus-0} the integration constants $M_{1,2}$ are naturally 
identified by orders of magnitude as in \rf{U-minus-magnitude}.
\item
Here the order of magnitude for $f_2$ is determined from the mass term of the
Higgs-like field $\s$ in the $(+)$ flat region resulting from \rf{U-plus-0} 
upon expansion around the Higgs vacuum 
($\s = \s_{\rm vac} + {\wti \s}$):
\be
\frac{{\widehat f}_1 m_0^2}{\chi_2 f_2}\, ({\wti \s}_a)^{*} ({\wti \s}_a) \; ,
\lab{Higgs-mass-term}
\ee
which implies that:
\be
f_2 \sim {\widehat f}_1 \sim M^4_{EW}
\lab{f2-scale-0}
\ee
in sharp distinction w.r.t. the order of magnitude of $f_2$ in \rf{f12-scales}
obtained within the formalism of Section 2.
\end{itemize}

The advantage of the formulation in this Appendix implementing Bekenstein's
idea about gravity-assisted spontaneous electro-weak symmetry breakdown over
the formulation in Section 2 of a slightly different mechanism for gravity-assisted
breaking versus restoration of electro-weak symmetry is that the
Einstein-frame effective action with the effective scalar potential 
\rf{U-eff-total-0} is renormalizable w.r.t. standard coupling-constant
renormalization procedure unlike the Einstein-frame action
\rf{einstein-frame-action} with the effective scalar potential \rf{U-eff-total}.
On the other hand, the formulation in Section 2 has the advantage of
yielding the value ($\sim 10^{-122} M^4_{Pl}$) of the vacuum energy density of 
the current (``late'') Universe directly in terms of the ``inflaton'' coupling
constants $f_{1,2}$ \rf{CC-eff-plus}, whereas in the Bekenstein-inspired
formulation in this Appendix we had to introduce {\em ad hoc} the ``late'' 
Universe vacuum energy density $\L_0$ as an independent free parameter.


\appendix
\section*{Appendix B}

Let us now consider briefly a slightly different version of the formalism in
Section 3 above. Namely, we can insist to incorporate the
Gauss-Bonnet-''inflaton'' coupling already from the very beginning within
the original generalized multi-measure gravity-matter action \rf{TMMT-1},
which will acquire the form:
\br
{\wti S} = \int d^4 x\,\P (A) \Bigl\lb g^{\m\n} R_{\m\n}(\G) + L_1 (\vp,X) + 
L_2 (\s,X_\s;\vp) \Bigr\rb
\nonu \\
+ \int d^4 x\,\P (B) \Bigl\lb U(\vp) + L_3 (\cA,\cB) +
\frac{\P (H)}{\sqrt{-g}}\Bigr\rb  + \int d^4 x \sqrt{-g}\,\vp \cR_{\rm GB} \; .
\lab{TMMT-2}
\er
Here $\cR_{\rm GB}$ is the standard Gauss-Bonnet scalar density in the {\em second
order} formalism w.r.t. original metric $g_{\m\n}$:
\be
\cR_{\rm GB} = R_{\m\n\kappa\lambda}(g) R^{\m\n\kappa\lambda}(g) - 4 R_{\m\n}(g) R^{\m\n}(g) + R^2 (g) \; .
\lab{GB-def-0}
\ee
A motivation to start with the action \rf{TMMT-2} is that it satisfies the
requirement for global Weyl-scale invariance under \rf{scale-transf}, which was 
crucial in order to fix uniquely the form of the initial multi-measure gravity-matter
action \rf{TMMT-1}.

Using the same steps as in Section 2 we arrive at the following
Einstein-frame action corresponding to \rf{TMMT-2}:
\br
{\wti S}_{\rm EF} = \int d^4 x \sqrt{-{\bar g}} \Bigl\lb R({\bar g}) + 
{\bar X} + {\bar X}_\s - \frac{1}{\chi_1} \Bigl(V_1 (\vp) + V_0 (\s)e^{\a\vp} + M_1\Bigr)
\nonu \\
+ \frac{\chi_2}{\chi^2_1}\Bigl(U(\vp) + M_2\Bigr)
- \frac{\chi_2}{4g^2} {\bar F}^2(\cA) - \frac{\chi_2}{4g^{\pr\,2}} {\bar F}^2(\cB)
- b\vp\(\frac{\cR_{\rm GB}}{\chi^2_1}\)_{g_{\m\n}=\chi^{-1}_1 {\bar g}_{\m\n}}\Bigr\rb
\; ,
\nonu \\
{}
\lab{einstein-frame-GB-0}
\er
where the last term explicitly reads (cf. \textsl{e.g.} \ct{dabrowski}; all objects 
on the r.h.s. are defined in terms of the Einstein-frame metric \rf{bar-g}):
\br
\(\frac{\cR_{\rm GB}}{\chi^2_1}\)_{g_{\m\n}=\chi^{-1}_1 {\bar g}_{\m\n}} =
{\bar \cR}_{\rm GB}
- 2 {\bar R}^{\m\n} \Bigl\lb 2 {\bar \nabla}_\m {\bar \nabla}_\n \ln\chi_1 +
\bigl({\bar \nabla}_\m \ln\chi_1\bigr) \bigl({\bar \nabla}_\n \ln\chi_1\bigr)\Bigr\rb
\nonu \\
+ 2 {\bar R}\bigl({\bar \Box} \ln\chi_1\bigr) 
+ 2\bigl\lb \bigl({\bar \Box} \ln\chi_1\bigr)^2  -
\bigl({\bar \nabla}^\m {\bar \nabla}^\n \ln\chi_1\bigr)
\bigl({\bar \nabla}_\m {\bar \nabla}_\n \ln\chi_1\bigr)\bigr\rb
\nonu \\
- \bigl({\bar \Box} \ln\chi_1\bigr) g^{\m\n}
\bigl({\bar \nabla}_\m \ln\chi_1\bigr) \bigl({\bar \nabla}_\n \ln\chi_1\bigr)
-2 \bigl({\bar \nabla}_\m {\bar \nabla}_\n \ln\chi_1\bigr)
\bigl({\bar \nabla}^\m \ln\chi_1\bigr) \bigl({\bar \nabla}^\n \ln\chi_1\bigr) \; ,
\nonu \\
{}
\lab{GB-EF}
\er
with ${\bar \cR}_{\rm GB}$ being the Einstein-frame Gauss-Bonnet density \rf{GB-def}.

We now observe a substantial physical difference between the Einstein-frame
theories \rf{einstein-frame-GB} and \rf{einstein-frame-GB-0}-\rf{GB-EF}.
In the latter case the field combination $\chi_1 = \P_1 (A)/\sqrt{-g}$
becomes an {\em additional physical propagating field degree of freedom} unlike in the
former case where it is just an algebraic function of the scalar matter fields 
\rf{chi1-eq}.

In particular, upon FLRW reduction 
(${\bar g}_{\m\n}dx^\m dx^\n = -N^2 (t) dt^2 + a(t) d{\vec x}^2$) 
the action \rf{einstein-frame-GB-0}-\rf{GB-EF} becomes (ignoring again  the
electro-weak gauge bosons, for simplicity):
\br
{\wti S}^{\rm (FLRW)} = \int dt \Bigl\{ - 6N a\adot + N a^3
\Bigl\lb \h \vpdot^2 +  (\nabla_0 \s)^{*}(\nabla_0 \s)
\nonu \\
- e^{-\psi}\Bigl(V_1 (\vp) + V_0 (\s)e^{\a\vp} + M_1\Bigr)
+ e^{-2\psi} \chi_2 \Bigl(U(\vp) + M_2\Bigr) 
\nonu \\
+b\,\vpdot \Bigl( 8\adot^3 - a^3 \psidot^3 + 6 \psidot^2 \adot a^2
-12 \psidot \adot^2 a\Bigr)\Bigr\rb \Bigr\} \; ,
\lab{FLRW-action}
\er
where:
\be
\psi \equiv \ln\chi_1 \quad ;\quad
\adot \equiv \frac{1}{N} \frac{da}{dt} \;\; ,\;\;
\vpdot \equiv \frac{1}{N} \frac{d\vp}{dt} \;\; ,\;\;
\psidot \equiv \frac{1}{N} \frac{d\psi}{dt} \; .
\lab{short-hand}
\ee

From the explicit form of \rf{einstein-frame-GB-0}-\rf{GB-EF} we deduce, that
corresponding equations for the extremums of the effective scalar field
potential (that is, for constant $\vp$, $\s$ and $\chi_1$) are {\em not affected}
by the presence of the additional terms in \rf{GB-EF} beyond the Einstein-frame
expression ${\bar \cR}_{\rm GB}$ \rf{GB-def} for the Gauss-Bonnet scalar
density, and they will reduce to Eqs.\rf{chi1-eq} and \rf{vp-vac-eq}-\rf{f1-tilde}.
However, when considering dynamical evolution -- for instance the Friedman
equations resulting from the FLRW action \rf{FLRW-action} -- there will be
an additional highly nonlinear evolution equation for the new dynamical
variable $\psi \equiv \ln\chi_1$ beyond \rf{friedman-1}-\rf{vp-eq-FRLW},
whose meaning is yet to be determined.



\end{document}